\documentclass[useAMS,usenatbib]{mnras}
\usepackage{epsfig}
\usepackage{threeparttable}
\usepackage{amsmath}
\title[Rapid orbital expansion in millisecond pulsar PSR J0636+5128]{Rapid orbital expansion in millisecond pulsar PSR J0636+5128: evaporation winds? }
\author[W. C. Chen]{ Wen-Cong Chen $^{1,2}$\thanks{E-mail:
chenwc@pku.edu.cn}\\
$^1$ School of Science, Qingdao University of Technology, Qingdao 266525, China;\\
$^2$ School of Physics and Electrical Information, Shangqiu Normal University,
Shangqiu 476000, China; \\
 }
\begin{document}

\date{}

\pagerange{\pageref{firstpage}--\pageref{lastpage}} \pubyear{2011}

\maketitle

\label{firstpage}

\begin{abstract}
PSR J0636+5128 is a 2.87 ms binary millisecond pulsar (MSP) discovered by the Green Bank
Northern Celestial Cap Pulsar Survey, and possesses the third shortest orbital period ($P=1.6$ hour) among confirmed binary pulsars. Recent observations reported that this source is experiencing a rapid orbital expansion at a rate of $\dot{P}=(1.89\pm0.05)\times 10^{-12}~\rm s\,s^{-1}$. The evaporation winds of the companion induced by the spin-down luminosity of the MSP may be responsible for such a positive orbital period derivative. However, our calculations show that the winds ejecting from the vicinity of the companion or the inner Lagrangian point can not account for the observation due to an implausible evaporation efficiency. Assuming that the evaporation winds eject from the vicinity of the MSP in the form of asymmetry disc winds or outflow, the evaporation efficiency can be constrained to be $\sim 0.1$. Therefore, the rapidly orbital expansion detected in PSR J0636+5128 provides evidence of outflows and accretion disc around the MSP.
\end{abstract}

\begin{keywords}
binaries: general --pulsars: general -- stars: evolution --  pulsars: individual (PSR J0636+5128)
\end{keywords}

\section{Introduction}
Millisecond pulsars (MSPs) are thought to be the evolutionary products of low-mass X-ray binaries (LMXBs), in which the neutron stars (NSs) accrete the material from the donor stars via the Roche lobe overflow. The NSs gain the angular momentum from the accreting material, and are spun up to millisecond period \citep{alpa82,bhat91}. Meanwhile, the NSs evolve into weak magnetic field ($10^{8}-10^{9}$ G) objects due to accretion-induced field decay \citep{kona97}. The final evolutionary products of LMXBs are sensitively related to the initial orbital periods. If the initial orbital periods are less than the so-called bifurcation periods, the LMXBs would evolve towards ultra-compact X-ray binaries (UCXBs) \citep{haaf12b,haaf12c,haaf13,chen16}, which are potential gravitational wave sources emitting low frequency signals \citep{taur18,chen20}. Since the binary MSPs have experienced a complicated mass transfer process, they are the ideal probes testing stellar and binary star evolutionary theory.

PSR J0636+5128 is a binary MSP discovered by the Green Bank
Northern Celestial Cap Pulsar Survey \citep{stov14}. This source includes a MSP with a spin period of 2.87 ms, and possesses the third shortest orbital period ($P=1.6$ hour) among confirmed binary pulsars. Its original name was PSR J0636+5129, while advanced
timing astrometry provides more accurate measurement, which yields an official name change \citep{drag18}. Assuming a MSP mass of $1.4~\rm M_{\odot}$, the minimum mass of the companion was derived to be $7\times 10^{-3}~\rm M_{\odot}$ by a small mass function.
Such properties including MSP, low-mass companion, tight orbit can classfy to "black widow" pulsar, in which a MSP accompanies by a low-mass ($\sim 0.01~\rm M_{\odot}$) companion in a tight orbit \citep{fruc88a,fruc88b,kulk88,stap96}. However, the absence of radio eclipses is not consistent with other black widow pulsars \citep{stov14}. This difference may arise from a relatively
face-on inclination \citep{drag18}. If the Roche-lobe filling, the minimum mean density of the companion was estimated to be about $43~\rm g\,cm^{-3}$ \citep{stov14}. By modeling the multicolor light curve, the mass and radius of the companion was derived to be $(1.71\pm0.23)\times10^{-2}~\rm M_{\odot}$ and $(7.6\pm1.4)\times10^{-2}~\rm R_{\odot}$ (assuming a pulsar mass of $1.4~\rm M_{\odot}$, Kaplan et al. 2018), respectively. Actually, a smaller filling factor yielded a higher mean density of $54\pm26~\rm g\,cm^{-3}$ \citep{kapl18}, which is obviously higher than the diamond planet PSR J1719-1438 \citep{bail11}.

According to 11-year time-of-arrival data by the \emph{North American Nanohertz Observatory for Gravitational Waves}, PSR J0636+5128 was experiencing an orbital expansion at a rate $\dot{P}=(2.5\pm0.3)\times 10^{-12}$ \citep{arzo18}. Recently, the observations by the \emph{Neutron star Interior Composition Explorer} refined the orbital period derivative to be $\dot{P}=(1.89\pm0.05)\times 10^{-12}$ \citep{guil19}. Kinetic effects caused by the relative kinematic motions of the binary MSP with respect to the solar system barycenter would induce an orbital period derivative \citep{shkl70,nice95}. However, \cite{kapl18} found that kinetic effects only contribute a fraction of 3\% in the observed orbital period derivative according to their estimation for the distance of this source. For a detached binary, gravitational radiation would cause the orbit to shrink. Therefore, there indeed exist other intriguing mechanism resulting in an expansion of the orbit. In this Letter we intend to diagnose whether or not the mass loss from this source could be responsible for the observed orbital period derivative. We investigate the orbital evolution of PSR J0636+5128 in section 2, and give a brief summary and discussion in section 3.

\section{Orbital evolution of PSR J0636+5128}

\subsection{Orbital period change of a binary}
PSR J0636+5128 includes a MSP and a low-mass companion, and its orbital angular momentum is
\begin{equation}
J=\frac{M_{\rm ns}M_{\rm c}}{M_{\rm ns}+M_{\rm c}}a^{2}\frac{2\pi}{P},
\end{equation}
where $a$ is the orbital separation, $P$ the orbital period, $M_{\rm ns}$, and $M_{\rm c}$ are the NS mass, and the companion mass, respectively. Differentiating equation (1), the change rate of the orbital period is given by \citep{chen19}
\begin{equation}
\frac{\dot{P}}{P}=3\frac{\dot{J}}{J}-3\frac{\dot{M}_{\rm c}}{M_{\rm c}}(1-q\beta)+\frac{\dot{M}_{\rm ns}+\dot{M}_{\rm c}}{M_{\rm ns}+M_{\rm c}},
\end{equation}
where $\beta=-\dot{M}_{\rm ns}/\dot{M}_{\rm c}$ is the accreting efficiency of the NS,  $q=M_{\rm c}/M_{\rm ns}$ is the mass ratio of the binary.

%The effective Roche-lobe radius of the companion star is \citep{egg83}
%\begin{equation}
%R_{\rm L}=\frac{0.49q^{2/3}}{0.6q^{2/3}+ {\rm
%ln}(1+q^{1/3})}a,
%\end{equation}

In principle, radio MSP has no mass accretion process, $\dot{M}_{\rm ns}=0$, and $\beta=0$, so equation (2) becomes
\begin{equation}
\frac{\dot{P}}{P}=3\frac{\dot{J}}{J}-3\frac{\dot{M}_{\rm c}}{M_{\rm c}}+\frac{\dot{M}_{\rm c}}{M_{\rm ns}+M_{\rm c}}.
\end{equation}
In a detached binary, gravitational wave radiation can result in an orbital shrinkage, and the orbital period derivative is
\begin{equation}
\dot{P}_{\rm gr}=-\frac{96G^{3}}{5c^{5}}\frac{M_{\rm ns}M_{\rm c}(M_{\rm ns}+M_{\rm c})}{a^{4}}P,
\end{equation}
where $G$ is the gravitational constant, $c$ the light velocity in
vacuo. Table 1 summarizes some observed and derived parameters of PSR J0636+5128. Taking $M_{\rm ns}=1.4~M_{\odot}$, $\dot{P}_{\rm gr}/P$ can be estimated to be $-7.4\times 10^{-18}~\rm s^{-1}$. In observation, $\dot{P}/P=3.3\times 10^{-16}~\rm s^{-1}$, so we ignore the contribution of gravitational radiation on the orbital evolution.

If the companion has a wind loss, the first and the third term on the right-hand side of equation (3) will contribute to negative period derivative. However, the second term should produce a positive period derivative.

\subsection{Winds ejecting from the vicinity of the companion}
When the MSP appears as a radio pulsar, its high energy radiation would evaporate the companion, and induce wind loss \citep{heuv88,rude89}. It is usually expected that the evaporation energy originate from the spin-down luminosity. \cite{kapl18} found that the photometric variability of the companion is dominated by the irradiation of the spin-down luminosity of the pulsar
in PSR J0636+5128, implying the possibility of the evaporation. The wind loss rate of the companion evaporating by the spin-down luminosity of the energetic MSP is given by \citep{heuv88,stev92}
\begin{equation}
\dot{M}_{\rm ev}=-f_{\rm ev}L_{\rm sd}\frac{R_{\rm c}^{3}}{4GM_{\rm c}a^{2}},
\end{equation}
where $f_{\rm ev}$ is the evaporation efficiency, and the spin-down luminosity
\begin{equation}
L_{\rm sd}=4\pi^{2}I\nu\dot{\nu},
\end{equation}
where $I$ is the moment of inertia of the NS, $\nu$ and $\dot{\nu}$ are the spin frequency and the spin frequency derivative, respectively. In the following calculations, we take $M_{\rm ns}=1.4~M_{\odot}$, and $I=10^{45}~\rm g\,cm^{2}$,

In this subsection, the evaporation winds is assumed to be ejected from the vicinity of the companion, carrying away the specific orbital angular momentum of the companion. Hence the angular momentum loss rate can be written as
\begin{equation}
\dot{J}_{\rm c}=\dot{M}_{\rm ev}\frac{M_{\rm ns}}{M_{\rm c}(M_{\rm ns}+M_{\rm c})}J.
\end{equation}
Inserting equation (7) into equation (3), the period derivative satisfies
\begin{equation}
\frac{\dot{P}_{\rm c}}{P}=-2\frac{\dot{M}_{\rm ev}}{M_{\rm ns}+M_{\rm c}}. \label{pdot-c}
\end{equation}
Actually, Eq. (\ref{pdot-c}) can also be derived by the Eq. (3) of \cite{salv08} when we take $\beta=0$ and $\alpha=1$ (where $\alpha$ is the specific-angular-momentum ratio between the outflowing material and the donor star).

From equations (5) and (8), we can obtain the evaporation efficiency as follows
\begin{equation}
f_{\rm ev}=\frac{\dot{P}}{P}\frac{2GM_{\rm c}(M_{\rm ns}+M_{\rm c})a^{2}}{L_{\rm sd}R_{\rm c}^{3}}.
\end{equation}
Adopting the parameters in Table 1, the evaporation efficiency $f=14.2$. It is implausible that the evaporation efficiency of the spin-down luminosity of the MSP exceeds $100\%$, hence the possibility that the evaporation winds eject from the vicinity of the companion can be ruled out.

\subsection{Winds ejecting from the inner Lagrangian point}
Similar to the Roche lobe overflow, the evaporation winds may penetrate into the Roche lobe of the NS through the inner Lagrangian point. However, the wind matter interacting with the strong radiation pressure of the pulsar will be accelerated and
ejected from the inner Lagrangian point during the radio-ejection phase of the MSP \citep{burd01,burd02,jia15}.

The distance between the mass center of the binary and the inner Lagrangian point is \citep{plav64,warn76}
\begin{equation}
b=(0.5-0.227{\rm log}q-\frac{q}{1+q})a. \label{b}
\end{equation}
Since $q\ll 1$ for PSR J0636+5128, so $b\approx(0.5-0.227{\rm log}q)a$. It is worth noting that Eq. (\ref{b}) will be invalid if the mass ratio $q<0.01$ (in this case, $b$ will exceed $a/(1+q)$, which represents the distance between the donor star and the mass center). For this source, $q\approx0.012$, we have $b\approx0.934a$, which is still in a valid range. Assuming that the outflowing matter carry away the specific angular momentum at the inner Lagrangian point, the angular momentum loss rate
\begin{equation}
\dot{J}_{\rm i}=\dot{M}_{\rm ev}(0.5-0.227{\rm log}q)^{2}\frac{(M_{\rm ns}+M_{\rm c})}{M_{\rm ns}M_{\rm c}}J.
\end{equation}
Due to $M_{\rm c}\ll M_{\rm ns}$, $\dot{J}_{\rm i}/\dot{J}_{\rm c}\approx(0.5-0.227{\rm log}q)^{2}=0.87$. Similar to the subsection 2.2, the evaporation efficiency can be estimated to be $f=0.84$. \cite{bret13} found that a fraction of $10-30\%$ energy received by the companion star from the spin-down luminosity of pulsars can be used to heat on the stellar surface, implying that the evaporation efficiency should be less than $0.1-0.3$. Therefore, it seems that the implausible evaporation efficiency can still not be solved even if the winds eject from the inner Lagrangian point.

\begin{table*}
\begin{center}
\centering \caption{ Some main observed and derived parameters for PSR J0636+5128.\label{tbl-1}}
%\begin{threeparttable}
\begin{tabular}{ccc}
\hline
Parameters & values & References\\
\hline
Spin frequency, $\nu$ (Hz) & 348.56   &1,2,3 \\
Spin frequency derivative , $\dot{\nu} (\rm s^{-2})$& $-4.19\times 10^{-16}$   &1,2,3 \\
Binary orbital period, $P$ (days)  & 0.06655  &1,2,3 \\
Binary orbital period derivative, $\dot{P}$ & $1.89(5)\times 10^{-12}$&2,3  \\
Companion mass, $M_{\rm c}(M_{\odot})$ & $1.71\pm0.23\times10^{-2}$& 4,5\\
Companion radius, $R_{\rm c}(R_{\odot})$& $7.6\pm1.4\times10^{-2}$ &4,5\\
\hline
\end{tabular}
   \begin{tablenotes}
     \item References. (1). \cite{stov14}, (2) \cite{arzo18}, (3) \cite{guil19}, (4) \cite{kapl18}, (5) \cite{drag18}.
    \end{tablenotes}
%\end{threeparttable}
\end{center}
\end{table*}

\subsection{Winds ejecting from the vicinity of the MSP}
It is inevitable to assume the winds ejecting from the vicinity of the MSP after the above two cases are unsuccessful. If the winds are transferred to the MSP, it will form an accretion disc around the MSP. The inner radius of the disc is generally thought to be at the magnetospheric radius, which can be derived by the equilibrium between the ram pressure of the infalling material and the magnetic pressure. The magnetospheric radius is \citep{davi73}
\begin{equation}
r_{\rm m}=\xi\left(\frac{\mu^{4}}{2GM_{\rm ns}\dot{M}^{2}}\right)^{1/7}\approx 2.1\times 10^{7}M_{\rm ns,1.4}^{-1/7}\dot{M}_{16}^{-2/7}\mu_{28}^{4/7}~\rm cm,
\end{equation}
where $\mu=BR^{3}/2=10^{28}{\rm G\,cm^{3}}\mu_{28}$ ($B$ is the surface dipole magnetic field of the NS) is the dipolar magnetic moment of the NS, $\dot{M}=10^{16}{\rm g\,s^{-1}}\dot{M}_{16}$ is the mass inflow rate in the disc, and we take the  dimensionless parameter $\xi=0.5$ \citep{ghos79}. The light cylinder radius of the MSP
\begin{equation}
r_{\rm lc} =\frac{c}{2\pi \nu}=1.4\times 10^{7}\nu_{\rm 348.56}^{-1}~\rm cm.
\end{equation}
If $r_{\rm m}>r_{\rm lc}$, the MSP will be detectable as an radio emission source.

During the radio phase, the inflow material will be ejected from the inner edge of the disc in the form of asymmetry disc winds or outflow, approximately carrying away the specific orbital angular momentum of the NS \citep{taur99}. In this case, the angular momentum loss rate of the winds
\begin{equation}
\dot{J}_{\rm n}=\dot{M}_{\rm ev}\frac{M_{\rm c}}{M_{\rm ns}(M_{\rm ns}+M_{\rm c})}J.
\end{equation}
Inserting equation (14) into equation (3), and taking into account $M_{\rm ns}\gg M_{\rm c}$, so the period derivative approximately obeys
\begin{equation}
\frac{\dot{P_{\rm n}}}{P}\approx-3\frac{\dot{M}_{\rm ev}}{M_{\rm c}}.
\end{equation}
According to equations (8) and (15), $\dot{P_{\rm n}}/\dot{P_{\rm c}}=\frac{3M_{\rm ns}}{2M_{\rm c}}$. Therefore, the period derivative that the winds eject from the vicinity of the MSP is about $120$ times that of the winds ejecting from the vicinity of the companion for PSR J0636+5128. The main reason causing this difference are as follows: the second term on the right-hand side of equation (3) should produce a positive period derivative, while the first and the third term would contribute to negative period derivatives; the distance between the NS and the mass center is obviously smaller that of the companion, hence it yields a low angular-momentum-loss rate and a small negative period derivative; for a specific second term with positive value, the first term with small absolute value naturally produces a high positive period derivative.

If the orbital expansion of PSR J0636+5128 fully arise from the evaporation winds ejecting from the vicinity of the MSP, the wind loss rate
\begin{equation}
\dot{M}_{\rm ev}=-\frac{M_{\rm c}\dot{P_{\rm n}}}{3P}\approx -3.7\times 10^{15}~\rm g\,s^{-1}.
\end{equation}
From equations (12), (13), and (16), and taking $r_{\rm m}>r_{\rm lc}$ into consideration, the magnetic field of PSR J0636+5128 can be constrained to be
\begin{equation}
B>3.0\times 10^{9}M_{\rm ns,1.4}^{1/4}\nu_{348.56}^{-7/4}~\rm G.
\end{equation}
This strength is slightly higher than normal magnetic field of the MSP population. If this source's spin-down energy is fully consumed by magnetic dipole radiation, the magnetic field can be estimated to be $B=3.2\times10^{19}\sqrt{-\dot{\nu}/\nu^{3}}/{\rm sin}\alpha~\rm G=1.0\times10^{8}~\rm G/{\rm sin}\alpha$, implying that the MSP may possess a small inclination angle with ${\rm sin}\alpha<0.03$.

According to equations (5), and (15), the evaporation efficiency can be expressed as
\begin{equation}
f_{\rm ev}=\frac{\dot{P}}{P}\frac{4GM_{\rm c}^{2}a^{2}}{3L_{\rm sd}R_{\rm c}^{3}}.
\end{equation}
We then estimate the evaporation efficiency to be $f_{\rm ev}=0.11$, which is in a plausible range.

\section{Summary and discussion}
In this Letter, we attempt to investigate the physical mechanism causing the orbital expansion of binary MSP PSR J0636+5128.
The observed period derivative is impossible to originate from gravitational radiation, which produces a contrary sign and low strength \citep{kapl18}. The pulsar wind of TeV ($e^{-},e^{+}$) particles could evaporate the companion of the MSP, driving an evaporation winds. It is generally thought that the evaporation winds eject from the vicinity of the companion or the inner Lagrangian point. However, our calculations indicate both cases can not account for the required amount for a plausible evaporation efficiency ($<1$). PSR J0636+5128 could provide a strong evidence that the evaporation winds eject from the vicinity of the MSP. Based on this assumption, the evaporation efficiency of PSR J0636+5128 is constrained to be $\sim0.1$. \cite{bret13} found an irradiation efficiency of $10-30\%$, which was defined as the effective fraction of the spin-down luminosity that the companion absorbed and re-radiated. If so, a fraction $10-15\%$ of the rest spin-down luminosity can be used to overcome the gravitation potential energy of the surface particles, driving evaporation winds.

The orbital expansion phenomenons similar to PSR J0636+5128 had already detected in some accreting MSPs such as SAX J1808.4-3658 \citep{salv08,hart08,hart09,burd09,patr12} and LMXBs such as 2A 1822-371 \citep{hell90,bayl10,burd10,iari11,chou16,mazz19}. A mass-loss rate of $10^{-9}~\rm M_{\odot}yr^{-1}$ from the donor star ejecting at the inner Lagrangian point may account for the observed period derivative of SAX J1808.4-3658 \citep{salv08,chen17}. Such a high mass loss rate from the donor star would require a relatively high evaporation efficiency of $30-40\%$ \citep{chen17,patr17}, or a large moment of inertia ($I\ga1.7\times 10^{45}~\rm g\,cm^{2}$) of the NS \citep{patr17}. Based on the model that the coupling between a relatively strong magnetic field of the donor star and the stellar winds induced by the X-ray radiation from the NS produces efficient angular momentum loss, \cite{xing19} successfully accounted for the fast orbital expansion, and high X-ray luminosity of 2A 1822-371. If the wind loss of the donor star eject from the vicinity of the MSP, a low mass loss rate would produce the observed period derivative without invoking high evaporation efficiency, large moment of inertia of the NS, or strong magnetic field of the donor star.

The orbital period and donor-star mass of PSR J0636+5128 are in good agreement with the mass-radius relations for low-mass white dwarfs given by \cite{delo03}. Therefore, it is possible that the progenitors of both PSR J0636+5128 and SAX J1808.4-3658 are UCXBs, in which a NS accretes the material from a white dwarf by the Roch lobe overflow \citep{haaf12a,haaf12b,seng17}. In principle, UCXBs firstly evolve to a minimum orbital period, and subsequently begin orbital expansion (see also Figure 2 in Chen \& Podsiadlowski 2016). Due to the decline of the mass transfer rate, the magnetospheric radius of PSR J0636+5128 penetrates outside the light cylinder, and the radio emission switches on \citep{li06}. After the NS is visible as a radio MSP, the evaporation winds induces the orbit to expand, and the companion decouples from its Roche lobe.

Actually, the orbital period also increases when the material is transferred from the less massive donor star to the more massive accretor. Considering a conservative mass and angular momentum transfer, $\dot{M}_{\rm ns}=-\dot{M}_{\rm c}$, and $\dot{J}=0$, equation (3) changes into
\begin{equation}
\frac{\dot{P}_{\rm tr}}{P}=-3\frac{\dot{M}_{\rm tr}}{M_{\rm c}}(1-q),
\end{equation}
where $\dot{M}_{\rm tr}$ is the mass transfer rate. Comparing with equation (15), the orbital period derivative of evaporation winds ejecting from the vicinity of the MSP is comparable with that of the mass transfer for $\dot{M}_{\rm tr}\approx\dot{M}_{\rm ev}$ and a small mass ratio. To yield the observed $\dot{P}$, it requires a mass transfer rate $\dot{M}_{\rm tr}\approx5.9\times10^{-11}~\rm M_{\odot} yr^{-1}$. However, the mass transfer rate in an UCXB is given by \citep{rapp87}
\begin{equation}
\dot{M}_{\rm tr}=6.21\times10^{-4}\left(\frac{M_{\rm ns}} {\rm M_{\odot}}\right)^{2/3}\left(\frac{P} {\rm minute}\right)^{-14/3}~\rm M_{\odot} yr^{-1}.
\end{equation}
Taking $M_{\rm ns}=1.4~\rm M_{\odot}$, PSR J0636+5128 only produces a mass transfer rate $\dot{M}_{\rm tr}\approx4.4\times10^{-13}~\rm M_{\odot} yr^{-1}$, which is two orders of magnitude smaller than the requirement value. Therefore, it is impossible for the mass transfer to yield the observed orbital period derivative.

The evaporation processes are also significant on the origin of black widows and redbacks. \cite{chen13} found that the determining factor that LMXBs evolve into either black widows or redbacks is the evaporation efficiency of the spin-down luminosity of the MSPs. The simulations of LMXBs evolution show that the binary systems without evaporation processes are difficult to evolve into black widows \citep{benv14}. \cite{smed15} proposed that evaporation process with $f_{\rm ev}\ga 0.12$ can ensure the companion to decouple its Roche lobe in the accretion-induced-collapse model. It strongly depend on the evaporation efficiency and the specific angular momentum extracting by the evaporation winds whether the redbacks evolve into black widows or MSP-He white dwarf binaries with wide orbits \citep{jia15}. PSR J0636+5128 is potentially related to the black widow MSPs \citep{stov14}, and also provides strong evidence that the evaporation winds eject from the vicinity of the MSP. Therefore, we propose that this evaporation winds model may be responsible for the origin of most black widows and redbacks.

There exist two promising observational checks whether the evaporation winds of PSR J0636+5128 eject from the vicinity of the MSP.
During the Roche lobe decoupling phase, the MSP would dissipate a fraction exceeding $50\%$ of the rotational energy \citep{taur12}. With the spin-down of the MSP, the light cylinder radius would increase to exceed the magnetospheric radius,
and the NS would appears as a X-ray source. If so, PSR J0636+5128 will have a chance to become a transitional MSP like PSR J1023+0038 \citep{arch09}, IGR J18245-2452 \citep{papi13}, and XSS J12270-4859 \citep{bass14}. In addition, if the NS hosts
an accretion disc with a radius extending up to the light cylinder radius, the emission in the UV/optical band should exist. However, somewhat faint optical spectroscopy of PSR J0636+5128 make difficult for this check channel \citep{kapl18}. Employing a U($344~\rm nm$) filter, PSR J0636+5128 was undetected in a total exposure time of 13.1 ks \citep{spie16}. Strong absorption of ejecta for UV emission may be responsible for undetection.

As an alternative mechanism, the orbital expansion observed in PSR J0636+5128 or SAX J1808.4-3658 could be also driven by the magnetic activities of the companions \citep{patr17}, which is called Applegate's mechanism \citep{appl92}. To account for the orbital period derivative of SAX J1808.4-3658, \cite{sann17} found that a surface magnetic field of $6000$ G is required to support a tidal dissipation mechanism acting on the donor star. Meanwhile, this scenario also hinted a great fraction of outflowing material that could obviously influence the orbital evolution of the binary. At present, it is uncertain whether such strong fields could exist in the companions of these two sources. Furthermore, it requires observations with baselines of decades to diagnose whether the gravitation quadrupole coupling can be responsible for the orbital evolution of PSR J0636+5128. We expect long-term multiwaveband observations for these sources can help us to untie the intriguing veil in the future.

\section*{Acknowledgments}
We thank the referee for a very careful reading and constructive comments that have led to the improvement of
the manuscript. We would also like to thank Xiang-Dong Li for helpful discussions. This work was partly supported by the National Natural Science Foundation of China (under grant number 11573016, 11733009), Program for Innovative Research Team (in Science and Technology) in University of Henan Province, and China Scholarship Council.\\
\\
\textbf{Data availability} \\
No new data were generated or analysed in support of this research.

\bsp

\label{lastpage}

\end{document}